\documentclass{aa}
\input psfig.sty

\begin{document}
%
%
   \title{The white dwarf cooling sequence in the old open cluster NGC\,188
\thanks{
Tables with the x, y coordinates, V and I magnitudes, are only 
available in electronic form at the CDS via anonymous ftp.}
}

   \subtitle{}

   \author{Andreuzzi Gloria \inst{1}, Richer Harvey B. \inst{2},  
Limongi Marco \inst{1}, Bolte Michael. \inst{3}}
   \offprints{Richer Harvey B.}
   \institute{Astronomical Observatory of Rome, via di Frascati I-33, 00040 
   Monteporzio Catone, Roma, Italy\\
   email: gloria@coma.mporzio.astro.it 
\and Department of Physics and Astronomy, University of British
Columbia, Vancouver, BC V6T 1Z1, Canada\\
\and University of California, UCO/Lick Observatory, Santa Cruz, CA 95064, USA  }
\date{Received / accepted}
\authorrunning{G. Andreuzzi et al.}
\titlerunning{White Dwarfs in NGC\,188}
 
\maketitle
 
\begin{abstract}
\label{sec:abstract}

We develop the white dwarf luminosity function (LF) of the old
open cluster NGC\,188 in order to determine a lower limit to the age
of the cluster by using the faint end of the cooling sequence.

To produce an extensive sequence of the cooling white dwarfs we imaged
four contiguous HST-WFPC2 fields in the center of the cluster in the
F555W and F814W filters.  After imposing selection criteria on the
detected objects we found a white dwarf cooling sequence (down to V
$\simeq$ 26.5) including 28 candidate white dwarfs in the cluster.  The
exposures are not deep enough to reach the end of this sequence, but
the results of our analysis allow us to establish a lower limit to the
age of the cluster independently of the isochrone fit to the cluster
turnoff.

The most ancient white dwarfs found are $\simeq$ 4 Gyr old, an age that
is set solely by the photometric limit of our data.  Classical methods
provide an estimate of $\simeq$ 7 Gyr (Sarajedini et al., 1999).

\end{abstract}

\section{Introduction}

Stellar evolution theory predicts that all single stars having a Main
Sequence (MS) mass
lower than $\simeq$ 8 M$_{\rm \odot}$ end their lives as white dwarfs
(WDs).  In
a star cluster, the white dwarf population carries key information for
addressing a number of astrophysical questions. In particular, 
since the white dwarfs cool at a predictable rate, their LF can be
used to constrain the cluster age. This age
estimate can provide an independent test of the more traditional age
determinations based on models and observations of the main-sequence
turnoff (MSTO).  

The white dwarf cooling sequence in the color-magnitude
diagram, in combination with models can be used to estimate the initial
mass -- final mass relationship between the stellar progenitors and the
white dwarfs over the range of progenitor mass from near
8M$_{\rm \odot}$ to the current-day turnoff mass of a cluster. This allows a
determination of the amount of mass lost from stars during their evolution
through stellar winds and planetary nebula ejection.

Ideally, to employ white dwarfs as age estimators, we need to observe a
large enough sample in any cluster to define the WDLF sufficiently well
to pinpoint the luminosity of the turndown with a precision equivalent
to a Gyr of cooling time. With smaller samples, it is still possible to
make some progress on the age questions. For example, even a single
good candidate WD in a cluster at magnitudes fainter than the faintest
WDs predicted from the MSTO age and cooling theory can bring into
question the vailidity of the WD {\it or} MSTO-based ages.

Clearly the most interesting case for testing MSTO-based ages is for
the globular clusters. However, because of the combination of large
age and large distance for even the nearest of the Galactic globulars,
it is not trivial to reach the end of the cooling sequence 
(e.g. Richer et al., 1995, 1997, 2002; Hansen et al. 2002). 
Old open clusters provide an interesting
alternative for WD-cooling age studies. There are several populous
clusters which are both nearer and younger than the nearest globulars
and for which the expected WDLF turndown is 3 magnitudes or more brighter.
For open clusters older than $\sim$ 4 Gyr, the main-sequence turnoff stars 
are very similar in structure to globular cluster turnoff stars, most 
importantly in that they have radiative cores, and comparison of nuclear and 
cooling ages for these old open clusters are very relevant the issue of 
globular cluster MSTO age tests. 
A significant difficulty with open clusters is that even the most populous
are sparse compared to most globular clusters. This leads to the problem that 
the lowest luminosity WDs are typically dramatically outnumbered by faint
blue galaxies in the cluster fields. Attempts to statistically
derive a WDLF by subtracting counts from nearby control fields are compromised
by small errors in control-field counts and cosmic dispersion in faint blue 
galaxy surface density. 
Very accurate star-galaxy separation is therefore crucial for
open cluster WD studies. Given the compact size of many faint blue galaxies, 
Hubble Space Telescope (HST) imaging gives a tremendous advantage.
Ultimately another powerful technique for isolating 
cluster-member WDs from the compact blue galaxy and field WD
populations may be measuring proper motions.

NGC\,188 was considered for a long time to be the oldest observable
open cluster in the Galaxy. Almost 40 years ago Sandage gave an
estimate for the age of this cluster of 10 Gyr (Sandage, 1962; see also
\cite{eggen}).  Recent papers revise the early age estimates
downward, and more modern values give an age near 7 Gyr (see
Sarajedini et al., 1999). Other open clusters have now been
found to be older than NGC\,188 (NGC\,6791 and Berkeley 17 with ages of
8 and 12 Gyr respectively, see \cite{phelps}) however, NGC\,188 is still one 
of the most studied since it is both nearer and less obscured than some of the
other old open clusters.  

Sects.  \ref{sec:obs} and \ref{sec:analysis} present the data, the 
reduction procedures and the color magnitude diagrams.
The resulting LFs together with final results are presented 
and discussed in  Sects.  \ref{sec:discussion} and \ref{sec:conclusion}.
 
\section{Observations and data reduction}
\label{sec:obs}

Deep HST WFPC2 images in  F555W and F814W were obtained around the centre 
 of NGC\,188 in the Cycle 7 program GO 7371.
The observations consisted of four fields, f1, f2, f3 and f4 located as shown 
in Table \ref{tab:jour}, in such a way that WF4 of the field f4 overlaps WF2 
of the field f1 and WF4 of the field f3 overlaps WF2 of the field f2.
Table \ref{tab:jour} also lists the corresponding exposure times together with
the number of observed frames for each filter. 

\begin{table}[ht]
\begin{small}
   \caption[]{Journal of observations}
   \label{tab:jour}
\[
\begin{array}{l|cccc}\cline{1-5}
{}   & {\rm \alpha}& {\rm \delta} & {\rm F555W} &  {\rm F814W}\\\cline{1-5}
{\rm f1}  & {\rm 0^h 48^m 00^s} & {\rm 85^0 13' 30''} &  4 {\rm \times} 1400 & 4 {\times} 1400    \\
{\rm f2}  & {\rm 0^h 48^m 00^s} & {\rm 85^0 16' 00''} &  4 {\rm \times}  1400 & 4 {\rm \times} 1400  \\
{\rm f3}  & {\rm 0^h 46^m 00^s} & {\rm 85^0 16' 00''} & 5 {\rm \times}  1400 & 4  {\rm \times}  1400   \\
{\rm f4} & {\rm 0^h 46^m 00^s} & {\rm 85^0 13' 30''} & 3  {\rm \times} 1300 &   3  {\rm \times} 1400     \\ \cline{1-5}
\end{array}
\]
\end{small}
\end{table}

Corrections to the raw data for bias, dark and flat-fielding were performed 
using the standard HST pipeline. 
Before reduction, we masked bad pixels and corrected the images for vignetting 
by using a mask and a vignetting mask respectively as built for HST data 
(supplied by P. Stetson).
We also multiplied our science exposures by a pixel-area map
to correct for geometrical distortion and to restore the integrity of the
flux measurements.
Photometry was measured using ALLFRAME (Stetson-version 3, 1997)
with a quadratically varying point spread function (PSF).
 
The images did not have enough bright unsaturated stars to create proper PSFs, 
so PSFs for each chip were constructed using WFPC2 images of the globular 
cluster $\omega$ Cen images (Stetson P. B.  1997, private communication).
Since there were virtually no stars in the PC frames, these were not
included in the reduction and analysis.
The data reduction process was completed by doing the charge transfer 
efficiency (CTE) correction on the data-set by using an algorithm supplied
by P. B. Stetson (priv. comm.). 

Finally the instrumental magnitudes were transformed into the standard 
Johnson system following \cite{holtzman}.

\begin{figure}[ht]
\psfig{figure=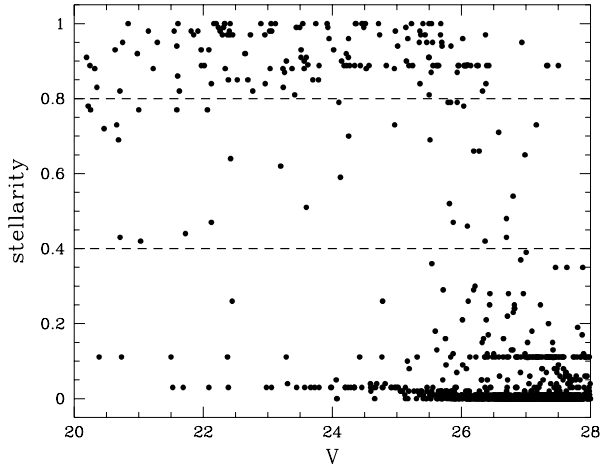,width=9truecm}
\caption{ \label{fig:stellarity} \baselineskip 0.4cm %
        SExtractor stellarity index {\it vs} the magnitude $V$ for objects in 
	our sample. 
	The horizontal dashed lines at $c = 0.4$ and $c = 0.8$ show
	the thresholds we used to distinguish between real stars and
	other objects (i.e. cosmic rays, image defects, galaxies).
	}
\end{figure}

To separate stars from other objects at faint magnitudes below V
$\simeq$ 24.5, we applied SExtractor to the images (\cite{bertin}).
This program uses a neural network to
distinguish between faint point sources from extended objects.
It assignes a stellarity index $c$ to the objects, ranging from 0 (galaxies) 
to 1 (stars).
Figure \ref{fig:stellarity} shows the results of the SExtractor
morphological classification versus the $V$ magnitude for all the objects in 
our sample. Looking at the figure there is a clear separation between the
objects with stellarity index around 1 and those with stellarity index
around 0 and there is very little in between.

Careful examination of each object by eye showed that virtually all the
objects for which 0 $\leq$ c $<$ 0.4 are classifiables as
galaxies, hot and warm pixels, ghost images or part of diffraction
spikes; all the objects for which  0.4 $\leq$ c $<$  0.8 
are dubious objects; all the objects for which c $\geq$ 0.8 
may be considered stars.

Using these criteria, we established a catalog of 157 stars in all the 
chips and all the fields studied selected from a total number of 1342 objects.

\section{Data analysis}
\label{sec:analysis}

\subsection{The colour-magnitude diagram and the WD cooling sequence}

\begin{figure}[ht]
\psfig{figure=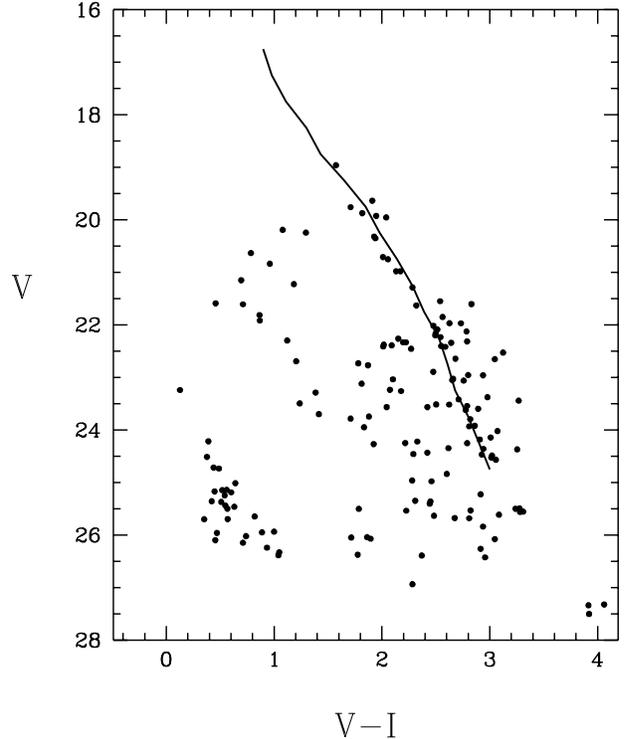,width=11truecm}
\caption{ \label{fig:cmd} \baselineskip 0.4cm %
$V-I$ colour-magnitude diagram for the objects in the four observed fields.
The solid line overlayed on the data is the ridge line obtained by using
the data sample from von Hippel et al. (1998).
}
\end{figure}

The $V$, $V-I$ CMD for all stars (as indicated from the stellarity index)
measured in the four fields is shown in 
Fig. \ref{fig:cmd}.

There is a fairly well-defined main sequence 
extending almost 6 magnitudes from V $\simeq$ 19 down to V $\simeq$ 25.
The solid line overlaid on our data is the ridge line obtained by using a 
sample of ground-based data (von Hippel et al., 1998). 

\begin{figure}[ht]
\psfig{figure=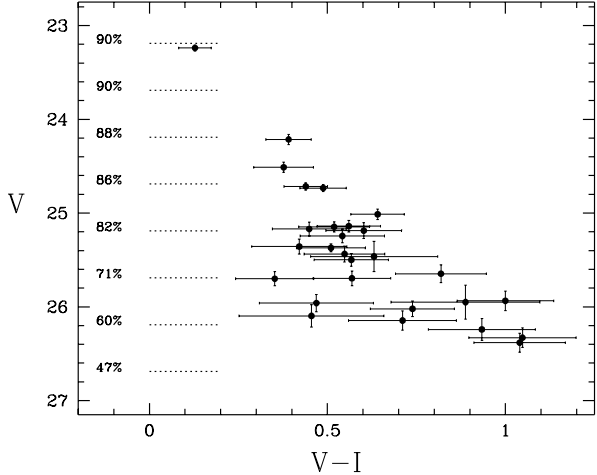,width=9truecm}
\caption{ \label{fig:cmdwds} \baselineskip 0.4cm %
         $V-I$ colour-magnitude diagram for all possible 
	 white dwarf candidates in our sample. Objects were
	 selected based on their stellarity index (c $\geq$ 0.8)
	 and, at this stage, there is no correction for field
	 contamination.
	 For each object we have also plotted the associated error
         based on the frame-to-frame scatter in the measurements.
	 Dotted lines (and the percentages) on the left 
	 indicate the mean completeness for each bin at the respective $V$ 
	magnitude.
	}
\end{figure}

In the lower left of the CMD there are 
28 blue objects ranging from V $\simeq$ 23 with V-I $\simeq$ 0  to V $\simeq$ 
26.5 with V-I $\simeq$ 1. 
An enlargement of this region of the CMD is shown in Fig. 
\ref{fig:cmdwds}; for each star we also plotted the associated photometric 
error as computed by ALLFRAME. 

Note that in this figure all objects appear to be stars but, for now,
we cannot discriminate between those belonging to the cluster and those
belonging to the Galactic field.  We expect a contamination of $\simeq$
8 Galactic white dwarfs at the magnitude and colour of interest for
our analysis (see Mendez \& Minniti, 2000).  The correction for the
field contamination will be addressed in the following sections.

We can use the CMD to estimate the expected number of
WDs in the cluster, starting from the number of the MS stars. 
To carry out this consistency check, we have found the slope of the
mass function of the cluster by using the number of MS stars in the V
magnitude range 20 -- 24, obtaining a value of 1.72 (Salpeter is
2.35).
We then estimated the mass of the stars at the TO under
the hypothesis that the cluster age is 4 -- 5 Gyr, obtaining respectively 
$\sim$ 1.23 M$_{\rm \odot}$ and 1.15 M$_{\rm \odot}$.
By looking at the coolest and faintest WDs inside
the range in magnitude V (23 -- 26), and adopting a WD mass of
0.6 M$_{\rm \odot}$, we obtained only 3 and 4 expected WDs in the cluster 
(respectively for 4 and 5 Gyr), a very small number if compared
with the observed number in the same range in magnitude (23 WDs).

We emphasize that this result is a consequence of 
the assumptions on the poorly determined mass function slope.  
For example, for a slope of 1.05, this number changes to 13 or 22
for cluster ages of 4 or 5 Gyr respectively. 
If we then are to believe that the most of the objects in the observed
cooling sequence are in fact WDs, we need to assume that our sample
is somehow deficient in MS stars. 
This is not unexpected for NGC\,188 because, as suggested by many
authors (Sarajedini et al., 1999, von Hippel et al., 1998), dynamical 
processes and mass segregation are important in the history of this
cluster which has the rather short relaxation time of t$_{\rm rh} = 6.4 
\times 10^7$ yr (McClure et al., 1977).

\subsection{Completeness correction}

Figure \ref{fig:cmdwds} also indicates the completeness in the counts
computed in bins of 0.5 magnitude in $V$.  Incompleteness corrections
were determined by carrying out artificial star tests on both sets of
the $V$ and $I$ frames, so as to be able to estimate the overall
completeness of our final colour magnitude diagram in the region of the
candidate white dwarfs.  To accomplish this we
first applied artificial star tests to the $I$ images.  Artificial stars
were added to random positions in the images over a range of 3
magnitudes in the region of the CMD of interest for the candidate white
dwarfs.  We created 5 artificial images for each original image each
with 100 artificial stars. We emphasize that crowding was not a problem
in our images.

We then added an equal number of stars at the same position of the
artificial $I$ stars to the $V$-band frames and with a magnitude such
that they would fall along the white dwarf cooling sequence.  All the
$V$ and $I$ frames with artificial stars were then subjected to the
same analysis adopted for the original frames.

An artificial star was considered recovered if $\delta{\rm X}$,
$\delta{\rm Y}$
$\leq$ 1.5 pixel and $\delta{\rm mag}$ $\leq$ 0.3 mag.  The ratio
$N_{\rm rec}/N_{\rm sim} = \Phi$, taken in 0.5 magnitude bins gives the
completeness in that bin for the image considered, where N$_{\rm rec}$ is
the number of the recovered stars and N$_{\rm sim}$ is the number of
simulated stars.

\subsection{Accounting for the field white dwarfs}

In addition to correction for photometric incompleteness
it is necessary to account for the contamination of the field stars.
The majority of the field star
contamination is due to main-sequence stars from the Galactic disk
together with a few main-sequence stars from the thick disk and the
halo. The majority of these objects have much redder colors than
our WD candidates in NGC\,188.

We have estimated the correction to our WD candidate counts,
by assuming that all resolved galaxies have already been eliminated 
from our data-sample. We then used the number of unresolved blue
objects identified by \cite{mendez} in the Hubble Deep Field 
North in an area $\simeq$ 4 times smaller than that covered by our
fields.

Table \ref{tab:lum} lists the counts for 0.5 magnitude-wide bins in $V$, 
the completeness factor and the counts corrected for this factor.
The last two columns report the number of unresolved blue objects
for each bin and our counts after the correction for completeness and
for the contribution of the field.\\
The absolute $V$ magnitudes of Col. 2 have been obtained adopting
(m-M)$_{\rm V}$ =
11.44, E$_{\rm B-V}$=0.09 (Sarajedini et al., 1999).

\begin{table}[ht]
$$\begin{array}{ll|lllcl}\cline{1-7}
{\rm V} & {\rm M_V} &{\rm N} & {\rm \Phi} & {\rm N_c} & {\rm N_{field}} & {\rm N_s}\\\cline{1-7}
23.19 & 11.75 & 1 & 0.9  & 1.11 & 0  & 1.11 \\
23.69 & 12.25 & 0 & 0.9  & 0.   & 0  & 0. \\
24.19 & 12.75 & 1 & 0.88 & 1.14 & 0  & 1.14 \\
24.69 & 13.25 & 3 & 0.86 & 3.49 & 0  & 3.49 \\
25.19 & 13.75 & 9& 0.82 & 10.97 & 0  & 10.97 \\
25.69 & 14.25 & 6 & 0.71 & 8.45 & 4  & 4.45 \\ 
26.19 & 14.75 & 8& 0.60 & 13.33 & 4  & 9.33 \\\cline{1-7}
\end{array}$$
\caption{\label{tab:lum}  \baselineskip 0.4cm %
          Luminosity function for the candidate white dwarfs
	  in our sample in the $V$ band in 
          0.5 mag bins. 
          N is the actual number of stars observed, $\Phi$ is the completeness 
          correction factor in each bin, N$_{\rm c}$ is the number of 
          stars after the 
	  corrections for incompleteness have been applied, N$_{\rm field}$ is
	 the number of unresolved blue objects as found in
	Mendez \& Minniti (2000) for the magnitude and the
	colour of our sample, N$_{\rm s}$ is the number of stars after
	the incompleteness correction and the field star subtraction,
	i.e. the cluster white dwarf luminosity function. 
}
\end{table}

\section{Discussion}
\label{sec:discussion}

\begin{figure}[ht]
\psfig{figure=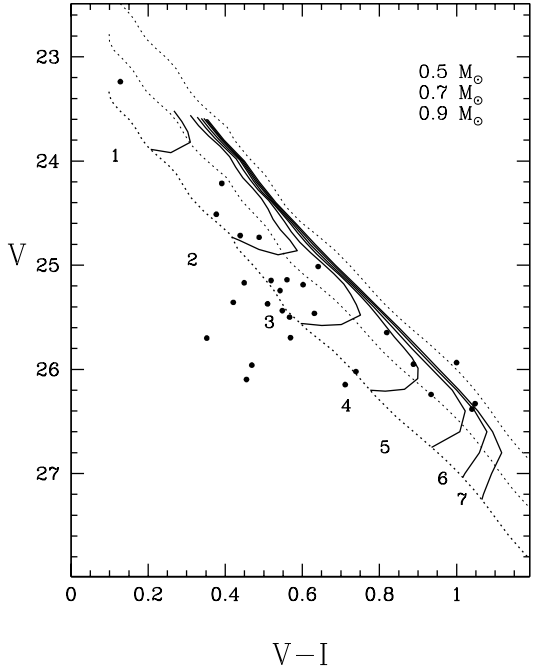,width=10truecm}
\caption{ \label{fig:iso} \baselineskip 0.4cm %
 Comparison of our NGC\,188 cooling sequence with theoretical
isochrones (solid lines) and hydrogen-rich WD cooling sequences (dashed
lines) obtained by using models from Hansen (1998, 1999).  
}
\end{figure}

To estimate the WD cooling age for NGC\,188, our first approach
is to compare the observed WD CMD with
the theoretical WD isochrones corresponding to different cluster
ages. 
Figure \ref{fig:iso} shows the theoretical WD isochrones corresponding to 
cluster ages of 1, 2, 3, 4, 5, 6 and 7 Gyr (solid lines), superimposed on
the observed NGC\,188 WD cooling sequence. In the figure we also plot the
theoretical cooling sequences for 0.5, 0.7 and 0.9  M$_{\rm \odot}$
hydrogen-rich WD (dotted lines) to indicate the mass range of the WDs
in the cluster. 
In particular most of the WDs cover the mass range included
in our models while a small fraction seem to have a mass larger than
0.9 M$_{\rm \odot}$.
Isochrones and theoretical cooling sequences have been
developed using the WD cooling models of Hansen (1998, 1999) as described in 
Richer et al. (2000) and have been transformed to apparent
magnitudes and colors appropriate for NGC\,188 adopting E(V-I) = 1.21 E(B-V) (based on
equations 3a and 3b of Cardelli et al., 1989 applied to the F814W
filter) and (m-M)$_{\rm V}$ = 11.44.

Since the population of WDs  accumulate at lower luminosities
as the age of the cluster increases, we can fix a limit to the age
of the cluster by simply looking at the location
of the WDs with respect to the theoretical isochrones.
Figure \ref{fig:iso} seems to show that for NGC\,188 this limit is $\sim$ 6
-- 7 Gyr.
However this method should be used with caution.
In fact, when we compare the observed WD CMD and the theoretical isochrones
we do not take into
account two important corrections that must be applied to
the data before carring out any quantitative conclusion, i.e., the correction for
incompleteness and the subtraction of the field contamination. These corrections
may be negligible when the sample is composed of many stars but they become
very important when we are dealing with such few objects. 

The second approach we can follow to derive an estimate of the age of
NGC\,188 is that of studying the observed WDLF.

As already mentioned above,
the population of WDs tend to accumulate at lower luminosities 
as the age of the cluster increases. Hence
the corresponding white dwarf LF will show a peak whose location
in magnitude is a function of the cluster age. In particular,
the peak of the white dwarf LF moves
towards fainter luminosities as the cluster ages (see Richer et al. 2000).

\begin{figure}[ht]
\psfig{figure=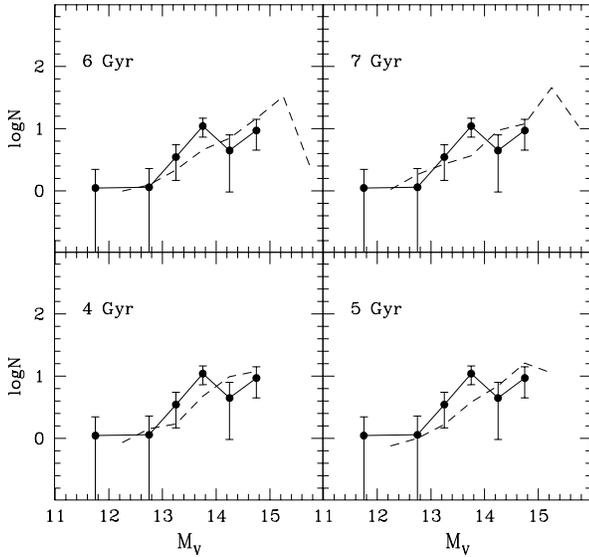,width=9truecm}
\caption{ \label{fig:diff} \baselineskip 0.4cm %
 Differential NGC\,188 white dwarf luminosity function (solid line),
 compared with synthetic luminosity functions of ages 
 4, 5, 6 and 7 Gyr (dashed lines).
}
\end{figure}

Figure \ref{fig:diff} (solid line) shows the differential observed
WDLF after corrections for incompleteness and field contamination
have been applied (see table \ref{tab:lum}). In the figure error bars
reflect the total error associated
with each bin and include both the counting uncertainty and the uncertainty
due to the correction for the incompleteness:

\begin{equation}
\sigma^2 \approx  \frac{\displaystyle N}{\displaystyle \Phi^2}
 + \frac{\displaystyle (1- \Phi) N^2} {\displaystyle N_{sim} \Phi^3}
\end{equation}

where N is the number of observed WDs; $\Phi$ 
is the incompleteness factor and N$_{\rm sim}$ is the number of
simulated stars.
 
The uncertainty in the number of objects for each bin has been estimated 
assuming that the counting uncertainies are derived from a Poisson 
distribution, and that the uncertainties in determining $\Phi$
are derived from a binomial distribution as shown in Bolte (1989).

\begin{equation}
\sigma^2_N = N
\end{equation}

\begin{equation}
\sigma^2_{\Phi} \approx \frac{\displaystyle \Phi (1- \Phi )} 
{\displaystyle N_{sim}}
\end{equation}

Dashed lines in the four panels of the figure refer to
4 synthetic LFs corresponding to 4, 5, 6 and 7 Gyr respectively.
They have been computed by adopting a Salpeter initial-mass function 
(Salpeter 1954, $n(m) \propto m^{- \alpha}; \alpha = 2.35$)
as described in Richer et al. (2000) and have been normalized to the total 
number of WDs brighter than 14.75, which is the magnitude of the 
faintest observed WD in the observed LF.

Figure \ref{fig:diff} shows that the peak of the theoretical LF
is located at magnitudes fainter than the last observed data point 
($M_{\rm V}$=14.75) only in the 6 and 7 Gyr cases.
Since it is unlikely that our data reach the true end of the 
cooling sequence, we suggest this is an indication that this cluster is at 
least as old as 5 Gyr.

\section{Conclusions}
\label{sec:conclusion}

Although we have identified a clear population of objects in
NGC\,188 with the photometric properties expected of cluster WDs, it
is apparent from Fig. \ref{fig:diff} that we have not reached the
turnover in the WDLF. The comparison with the LF models suggests that
we can place a lower limit of 5 Gyr on the age of the cluster.

These observations could be improved significantly with a modest
effort, with HST and the ACS. A fainter photometric limit together with an
expanded field coverage would allow a large sample of WDs to be
identified in the NGC\,188 field. 
NGC\,188 stars have a proper motion of 2.3 milli-arcseconds with
respect to the field and with a followup program, a subsample of
bona-fide cluster WDs could be identified on the basis of their proper 
motions.

\begin{acknowledgements}
We warmly thank P. B. Stetson for kindly providing us images and files for
data reduction.
We also warmly thank P. B. Stetson, G. Marconi and A. Chieffi, for useful
discussions, comments and suggestions.
Gloria Andreuzzi gratefully acknowledges the hospitality of H. Richer
and the University of British Coloumbia. 
This work has been supported by MURST/Cofin1999 under the project: 
``Effect of the dynamics on the canonical and exotic stellar distributions 
inside galactic globular clusters''.
\end{acknowledgements}

\end{document}